--------------------------------------------------------------------------
\documentstyle[aps,preprint]{revtex}
\begin{document}
\draft
\title{ Topological spin excitations of
Heisenberg antiferromagnets in two dimensions
}
\author{Tai Kai Ng}
\address{Dept. of Physics, HKUST, Kowloon, Hong Kong}
\date{ \today }
\maketitle
\begin{abstract}
  In this paper we discuss the construction and the dynamics of 
vortex-like topological spin excitations in the
Schwinger-boson description of Heisenberg
antiferromagnets in two dimensions. The topological spin
excitations are Dirac fermions (with gap) when spin value $S$ 
is a half-integer. Experimental and
theoretical implications of these excitations
are being investigated.
\end{abstract} 


\narrowtext
   Recently there are strong interests in the study of Heisenberg
antiferromagnets on two dimensional square lattice. Early theoretical
analysis showed that the low energy, long length scale behaviour
of the model should be described by the renormalized classical
regime of the $(2+1)d$ O(3) non-linear-sigma-model 
(NL$\sigma$M)\cite{chn,man}. The analysis was believed to be valid
as long as the length scale under investigation is much
larger than lattice spacing.
More recently, in a series of experiments\cite{e1} and Monte
Carlo simulations\cite{mc1,mc2,mc3}, it was discovered that the
length scale at which the renormalized classical $\sigma$-model
description becomes valid is surprisingly long ($L\geq200$ sites
for spin $S=1/2$)\cite{mc2,mc3}. Quantum Monte Carlo study of the
low energy spectrum of the model on finite size lattices
also suggested that the spectrum in the $S=1/2$ case disagrees
rather strongly with prediction of NL$\sigma$M even when the
size of the system $N$ is not too small ($N\leq32$)\cite{mc4}.
These anomalous findings suggest that contrary to usual believe,
there may exist an intermediate energy/length scale where
the behaviours of the system deviates strongly from
NL$\sigma$M. 

  On the other hand, it has been shown by Read and 
Chakraborty\cite{rc} that fermionic spin excitations exist as
topological excitations in the short-ranged RVB phase of 2D
$S=1/2$ Heisenberg model, and their analysis has been generalized
by the author\cite{ng} to the case (for arbitrary spin $S$) when
the system is described by Schwinger-boson mean-field
theory(SBMFT)\cite{aa}. These topological excitations are finite
energy excitations of 2D Heisenberg antiferromagnets that
do not appear in conventional NL$\sigma$M description.
However, the dynamics of the topological spin excitations
were not investigated in those works\cite{rc,ng}. 

    The purpose of the present paper is to investigate, starting
from SBMFT, the dynamics of the topological spin excitations
constructed in Ref.\cite{ng} under some very general assumptions. 
In particular, we shall show that the existence of these
topological excitations is consistent with the
recently observed anomalous behaviours in the Heisenberg model.
We shall also point out, in the case of $S=1/2$ Heisenberg
antiferromagnet, an interesting connection between the
topological spin excitations and the spin excitations in the
flux-phase\cite{ma} of fermionic RVB mean-field theory.

   To start with we review briefly the construction of topological
spin excitations in SBMFT of 2D Heisenberg model.
Following Ref.\cite{ng} we assume that the low energy physics
of general spin-$S$ 2D Heisenberg model is described
by the Schwinger-boson mean-field Hamiltonian
\begin{equation}
\label{hmf}
H_{MF}=-J\sum_{i,j}\left(\Delta^*(Z_{i\uparrow}Z_{j\downarrow}
-Z_{i\downarrow}Z_{j\uparrow})+H.C.-|\Delta|^2\right)
+\sum_{i\sigma}\lambda(\bar{Z}_{i\sigma}Z_{i\sigma}-2S),
\end{equation}
where $\Delta(\Delta^*)$ and $\lambda$ are mean-field parameters
determined by the mean-field equations $\Delta=<(Z_{i\uparrow}
Z_{j\downarrow}-Z_{i\downarrow}Z_{j\uparrow})>$ and $<\bar{Z}_{i\uparrow}
Z_{i\uparrow}>+<\bar{Z}_{i\downarrow}Z_{i\downarrow}>=2S$.
Notice that in two dimensions, long-ranged antiferromagnetic order
exists at zero temperature corresponding to bose-condensation
$<Z>\neq0$ in SBMFT. We shall consider
finite temperature $T\neq0$ and $<Z>=0$ in the following. The
effect of bose condensation will be addressed at the end of the paper.
Notice that SBMFT offers a fairly accurate 
description of the low energy physics of Heisenberg model at two 
dimensions\cite{aa}, thus justifying the starting point of our theory.

   To look for topological excitations in SBMFT, we notice that
the structure of the mean-field theory resembles very much the BCS
theory for superconductivity, except that the spin pairs of bosons
replace the electron (fermion) Cooper pairs in BCS theory. 
The resemblance of the two theories leads
us to study {\em vortex} excitations in SBMFT, since
vortices are stable topological excitations in BCS theory at
two dimensions. In BCS theory, a vortex located at $\vec{r}=0$
is a solution of the BCS mean-field equation where the order
parameter $\Delta_{BCS}(\vec{r})$ has a form 
$\Delta_{BCS}(\vec{r})=f(r)e^{i\theta}$
in polar coordinate, where $f(r)$ is real and positive. To
minimize energy, a magnetix flux of $\pi$-flux quanta is trapped
in the vortex core. The vortex solution
in SBMFT has the same structure, except that the BCS
order parameter $\Delta_{BCS}$ is replaced by the Schwinger
boson order parameter $\Delta_{ij}$ and the vector potential 
$\vec{A}$ does not represent the physical magnetic field,
but is a fictitious gauge field arising from phase fluctuations
of order parameter $\Delta_{ij}$'s\cite{ng,rs}. The
existence and stability of the vortex solution
was demonstrated in Ref.\cite{ng} in an effective
Ginsburg-Landau theory. These vortex solutions
are bosonic, $S=0$ topological excitations in SBMFT\cite{ng}.

  To construct topological {\em spin} excitations we note that like
vortices in superconductors where electronic bound states often exist
inside the vortex core, boson bound states may exist inside 
vortices in SBMFT. In particular, we argued in Ref.\cite{ng} 
that for vortex centered at a lattice site, a bound 
state of $2S$ bosons must be formed at the vortex center because
of the constraint that there is always $2S$ bosons per site in the
Heisenberg model. In particular, because of statistics transmutation
associated with binding quantum particles to flux-tube of $\pi$-flux
quantum in two dimensions\cite{wz}, the resulting 
excitation is a spin-$S$ fermion when $2S$ is odd\cite{ng}. 

 We shall now supply the mathematical details. An important
difference between SBMFT and BCS theory is that the mean-field 
Hamiltonian \ (\ref{hmf}) breaks translational symmetry of the
Heisenberg model by one lattice site. Correspondingly there exists
two lattice sites per unit cell in SBMFT and the
fluctuations of the order parameter $\Delta_{ij}$'s are described
by two amplitude and two phase (uniform and staggered) fields in 
the continuum limit\cite{ng,rs}, i.e.
\begin{equation}
\label{delta}
\Delta_{i,i\pm\nu}={1\over2}\left(\phi(i\pm{\nu\over2})+
q_{\pm\nu}(i\pm{\nu\over2})\right)e^{i\left[\int^{i\pm{\nu\over2}}
\vec{A}.d\vec{x}+A^s_{\pm\nu}(i\pm{\nu\over2})\right]},
\end{equation}
where $q_{-\nu}=-q_{\nu}$, $A^s_{-\nu}=-A^s_{\nu}$ are 'staggered' 
components of the amplitude and phase fluctuations of $\Delta$, 
respectively. $\phi$ and $\int^x\vec{A}.d\vec{x}'$ are the corresponding 
'uniform' components. The effective Ginsburg-Landau action for the 
continuum field variables is derived in Ref.\cite{ng}(see also
Ref.\cite{rs}). We obtain to order $O(m^0)$ ($m$ is the mass gap for
spinwave excitations in SBMFT which is very small at low temperature
($m\sim{J}e^{-JS/T}$)),
$S_{eff}=S_{u}+S_{s}$, where
\begin{eqnarray}
\label{gl}
S_{u} & \sim & \int{d}\tau\int{d}^2x\left((a-4(S+{1\over2}))\phi+\phi^2
+(S+{1\over2})\phi(\partial_{\mu}\theta-A_{\mu})^2\right),
\\ \nonumber
S_{s} & \sim & \int{d}\tau\int{d}^2x\left(({1\over2}-{b\over\phi})(q_{\mu})^2
+{1\over2e^2}F_{\mu\nu}^2+icF_{\mu\tau}q_{\mu}\right),
\end{eqnarray}
where $e^2\sim{m}$, $a<4$, $b$ and $c$ are constants of order O(1).
$F_{\mu\nu}=\partial_{\mu}A^s_{\nu}-\partial_{\nu}A^s_{\mu}$ and
$\theta/2$ is the "uniform" phase field for the Schwinger bosons.
Notice that the gauge symmetry of the 'uniform' gauge field
$\vec{A}$ is broken by $<\Delta_{ij}>=\phi\neq0$ in SBMFT. 
The existence of stable vortex solution in SBMFT is tied with the
$\vec{A}$ field as in usual superconductors.

  To derive vortex dynamics we concentrate at $S_{u}$ and
introduce vortex 3-current $\vec{j}^v$ in the phase field,
$\partial_{\mu}\theta\rightarrow\partial_{\mu}\theta
+\partial_{\mu}\theta'$, where $\theta'$ is multivalued and $j^v_{\mu}
=\epsilon_{\mu\nu\lambda}\partial_{\nu}\partial_{\lambda}\theta'\neq0$.
In the London limit where we treat $\phi$ as constant a duality
transformation can be performed where we can integrate out the
$\theta$ field to obtain an effective action $S_{v}$ for
vortices\cite{fl},
\begin{mathletters}
\label{sv}
\begin{equation}
\label{sv1}
S_{v}=\int{d}\tau\int{d}^2x\left({1\over4\phi}(\nabla\times\vec{a})^2
+i\vec{a}.(\vec{j}^v-\nabla\times\vec{A})\right)+S^{(0)}_v
\end{equation}
where $\nabla\times\vec{a}\sim(\nabla\theta'-\vec{A})$ and corresponds
to the transverse part of supercurrent field in superconductors.
Notice that the amplitude field $\phi$ goes to zero at the center of
vortex. As a result there exists an additional contribution to the
vortex action, $S^{(0)}_v\sim\{$(energy needed
to create vortex core, $\epsilon_v$)$\times$(length of vortex 
trajectory in space-time)$\}$. For $N$ vortices,
\begin{equation}
\label{sv2}
S^{(0)}_v\sim{\epsilon}_v\sum_{i=1}^N\int{d}l_i=\epsilon_v
\sum_{i=1}^N\int{d}\tau\sqrt{1+{1\over{c_v}^2}
({d\vec{x_i}\over{d}\tau})^2},
\end{equation}
\end{mathletters}
where $\vec{x}_i(\tau)$ represents the trajectory of the $i$th vortex
in Euclidean space-time, $c_v$ is the limiting velocity. Notice
that the suppression of $\phi$ field at vortex core also couples
vortices to the staggered amplitude and phase fluctuations through
the term $(-{b\over\phi})q_{\mu}^2$ in $S_s$\cite{ng}. We shall
first consider $S_v$ in the following.

  The dynamics of a single vortex can be obtained by minimizing  
$S^{(0)}_v$ at real time. We find that $S^{(0)}_v$ describes
relativistic particles with energy $E=\gamma{\epsilon_v}$, where
$\gamma=1/\sqrt{1-{\vec{v}^2\over{c}_v^2}}$ and $\vec{v}$ is the
vortex velocity. In the absence of $\nabla\times\vec{A}$ term the
particles have 'charge' (vorticity) and interact with each other through
an effective U(1) gauge field $\vec{a}$. For boson vortices 
the corresponding quantum field theory is a relativistic theory of
scalar electrodynamics with charged bosons (vortices)\cite{fl}.
In the presence of trapped magnetic flux inside vortex core
$(\nabla\times\vec{A}-\vec{j}_v\sim0)$, the electric 3-current
$\vec{j}_v$ is screened and the bosons decoupled from the
gauge field $\vec{a}$. The resulting theory is a
relativistic theory of bosons with short-ranged interactions,
as is in the case of vortices in usual superconductors\cite{fl}.

  To derive dynamics for the topological spin excitations we assume that 
once the boson spins are bound to the vortex core, their spatial
degree of freedom is quenched and the only modifications to the
pure vortex action \ (\ref{sv}) are: (1)the vortices now
carry spin indices $m=-S,-S+1,...,S$, and there are $2S+1$ spin
component of vortices, and (2)vortices become fermions when
$2S$ is odd. In particular, since the vortex 
action \ (\ref{sv}) is Lorentz invariant, the
dynamics of the topological spin excitations
must be described by $(2+1)d$-relativistic field theories of 
bosons when $S$ is integer and $(2+1)d$-relativistic field
theories of fermions when $S$ is half-integer. To proceed further
we examine more carefully the symmetry constraints in SBMFT.
The existence of two lattice sites per unit cell in SBMFT
implies that there are also two quantum fields $\psi^{A}
_{\sigma}$ and $\psi^{B}_{\sigma}$ in the continuum theory,
representing spin-$\sigma$ vortices centered at $A$-
and $B$- sublattice sites, respectively\cite{ng}. Under reflection
or rotation by $\pi/2$ around center of a square plaquette, the 
$A$- and $B$- sublattices are interchanged and correspondingly
also $\psi^{A}_{\sigma}$ and $\psi^{B}_{\sigma}$. Notice that
we have considered finite temperature where $<Z>=0$ in our discussion
and correspondingly parity (space-time reflection) symmetry
is unbroken.
In order to describe coupling of the bound boson
spins at the vortex center to the fictitious
gauge fields $\vec{A}$ and $\vec{A}_s$, the
quantum fields $\psi^{A}_{\sigma}$ and $\psi^{B}_{\sigma}$
must also carry 'charge' and are complex. 

 With these kinematics constraints, the quantum field theories
for the topological spin excitations can be determined\cite{qft}.
In the case of integer spins where the topological spin excitations
are bosons, we construct quantum fields $\psi^{+(-)}_{\sigma}=\psi^{A}
_{\sigma}\pm\psi^{B}_{\sigma}$ which are eigenstates of
$\pi/2$ rotation and reflection with eigenvalues $\pm1$.  
The dynamics of the $\psi^{+(-)}_{\sigma}$ fields are described
separately by relativistic theories of complex 
scalar-fields. In the case of half-integer spins where
the spin excitations are fermions the corresponding theory
which respects parity is a theory
of Dirac fermions in $(2+1)d$\cite{qft}. In the  
following we shall concentrate at the case of half-integer
spin systems.

In this case, the topological spin excitations are Dirac spinors
described by four-component spinor field $\psi_{\sigma}$, with
\begin{equation}
\label{dspinor}
\psi_{\sigma}(x)=\left(\begin{array}{r}
\psi^{A}_{\sigma}(x)  \\
\psi^{B}_{\sigma}(x)
\end{array}\right)\; , \;\;\;
\psi^{A(B)}_{\sigma}(x)=\left(\begin{array}{r}
\psi^{A(B)}_{1\sigma}(x) \\  
\psi^{A(B)}_{2\sigma}(x)  
\end{array}\right)
\end{equation}
where $\psi^{A(B)}_{\sigma}$'s are two component fermion 
fields needed to describe positive and negative energy 
solutions of the Dirac equation. In terms of $\psi_{\sigma}$
the effective Lagrangian which transforms correctly under
parity is
\begin{equation}
\label{leff}
L_{eff}=\sum_{\sigma}{i\over2}[\bar{\psi}_{\sigma}\gamma^{\mu}
(\partial_{\mu}\psi_{\sigma})-(\partial_{\mu}\bar{\psi}_{\sigma})
\gamma^{\mu}\psi_{\sigma}]-m\bar{\psi}_{\sigma}
\psi_{\sigma}-\bar{\psi}_{\sigma}\gamma^{\mu}\psi_{\sigma}A_{\mu},
\end{equation}
where $\gamma^{\mu}$'s are usual $4\times4$ Dirac
matrices in $(2+1)d$ with $\mu=0,1,2$. $\vec{A}$ is the 'uniform'
gauge field. Notice that the
staggered gauge field $\vec{A}_s$ decouples from
the $\psi_{\sigma}$ field in this level, as can be checked easily
from transformation of $L_{eff}$ under staggered gauge transformation
$\psi^{A}_{\sigma}\rightarrow\psi^{A}_{\sigma}e^{i\theta}$, and
$\psi^{B}_{\sigma}\rightarrow\psi^{B}_{\sigma}e^{-i\theta}$.
Notice also that the spin degrees of freedom $\sigma$
appear as {\em internal} degrees of freedom for the 
Dirac fermions.

  It is interesting to compare the effective action $L_{eff}$
with the flux-phase\cite{ma} of the fermionic mean-field theory
of Heisenberg model in the case
$S=1/2$. In both cases pockets of Dirac fermions describe the 
fermionic spin excitations at low energy. The
number of species of Dirac fermions are the same in both cases -
there are four 'half-pockets' of Dirac-fermions
around $(k_x,k_y)=(\pm{\pi\over2},\pm{\pi\over2})$ in
the flux-phase, and there are two 'full-pockets' of Dirac
fermions coming from two independent combinations of
$\psi^{A}_{\sigma}$ and $\psi^{B}_{\sigma}$ in the present theory. 
The position of the 'Dirac-pockets' in $\vec{k}$-space cannot be
determined with certainty in our effective theory. Nevertheless,
to make comparison with the flux-phase
we shall assume that the fermion pockets in our
theory are centered around $(k_x,k_y)=(\pm{\pi\over2},
\pm{\pi\over2})$. With this assumption the only difference between
$L_{eff}$ and the flux-phase is that there is a gap $\epsilon_v$
in the Dirac-fermion spectrum in our theory. In fact, in the limit
$\epsilon_v=0$ it can be checked directly that $L_{eff}$ describes
the continuum theory for the flux phase. The strong
similarity between our effective theory and flux-phase leads 
us to speculate that the flux-phase in fact describes 
a new spin-disordered phase of 2D Heisenberg
antiferromagnets where antiferromagnetism is destroyed
by driving the fermion mass-gap $\epsilon_v$ to zero.
We shall discuss this scenario in a separate paper.

 Next we consider the experimental consequences of the
topological spin excitations. We shall first consider $S=1/2$ as 
example. In terms of
$\psi$, The spins carry by the Dirac fermions in the $S=1/2$ case
are described by the operators
\begin{equation}
\vec{S}^{A(B)}(x)={1\over2}\sum_{m,m'}\sum_{k=1,2}
\left(:\psi^{+A(B)}_{km}(x)
\vec{\sigma}_{mm'}\psi^{A(B)}_{km'}(x):\right)
\label{spino}
\end{equation}
where $\vec{S}^A(x)$ and $\vec{S}^B(x)$ are spin operators in the
$A$- and $B$- sublattices, respectively and $\vec{\sigma}$ are
Pauli matrices, $m,m'=\pm{1\over2}$. $:\hat{O}:=\hat{O} 
-<\hat{O}>_G$ are normal-ordered operators. The
'uniform' and 'staggered' spin operators can be constructed from
$\vec{S}^{A(B)}$, where $\vec{m}(\vec{n})=\vec{S}^A+(-)\vec{S}^B$ 
describes spin fluctuations around momenta $\vec{q}=(0,0)$ and
$\vec{q}=(\pi,\pi)$. Notice that
because of the constraint that there is one spin per site
in the original Heisenberg model, the Dirac fermions 
together with the original boson spins must satisfy the constraint
\begin{equation}
\label{constraint}
\sum_{\sigma}\left((\sum_{k=1,2}:\psi^{+\alpha}_{k\sigma}(x)
\psi^{\alpha}_{k\sigma}(x):)
+\bar{Z}^{\alpha}_{\sigma}(x)Z^{\alpha}_{\sigma}(x)\right)=1,
\end{equation}
which introduces additional coupling between boson spins
and topological spin excitations not included in $L_{eff}$.

  Experimentally the Dirac fermion spectrum can be observed directly
in neutron scattering experiment at energy $\omega>2\epsilon_v$.
Their contribution to dynamic structure factor $S(\vec{q},\omega)$ 
at $\vec{q}\sim(0,0)$ and $\vec{q}\sim(\pi,\pi)$ can be calculated 
directly using Eqs. \ (\ref{leff}) and \ (\ref{spino}). We find that
their contributions are similar to $S(\vec{q},\omega)$ calculated
from flux phase, except that a gap $\sim2\epsilon_v$ is found in the 
spectral function. The existence of Dirac fermion spectrum also
affects spin correlation at lower temperatures. To see that
we consider the antiferromagnetic spin-correlation length $\xi(T)$.
In pure SBMFT, $\xi(T)$ is given at low temperature by\cite{aa}
\begin{equation}
\label{xi}
\xi(T)=\sqrt{2}{\Delta{J}\over{T}}exp({2\pi\Delta{J}\over{T}}n_B), 
\end{equation}
where $n_B$ is the density of bose-condensed spins in SBMFT
at $T=0$. In the presence of Dirac fermion spectrum,
the system gets more disordered at $T\neq0$
because of thermal effect associated with the extra degrees 
of freedom and $\xi(T)$ decreases. The effect can be estimated
in a mean-field approximation using the constraint \ (\ref{constraint})
where the average number of Schwinger boson per site $<n_{SB}>$ is
given by
\[
<n_{SB}>=1-\sum_{k=1,2,\sigma}<:\psi^{+\alpha}_{k\sigma}\psi^{\alpha}
_{k\sigma}:>.  \]
  It is straightforward to show that at low temperature, 
$<n_{SB}>\sim1-2exp(-{\epsilon_v\over{T}})$, and the leading correction
at low temperature to $\xi(T)$ is obtained by replacing $n_B\rightarrow
{n}_B-2exp(-{\epsilon_v\over{T}})$ in Eq.\ (\ref{xi}). The reduction in
$\xi(T)$ continues until $T\sim\epsilon_v$, when
the number of thermally excited fermion spin excitations
becomes large, and the spin correlation becomes qualitatively 
different from the prediction of SBMFT or NL$\sigma$M. For
general spin value $S$, $\epsilon_v\sim{J}S^2$ and the temperature
at which the NL$\sigma$M description becomes invalid occurs at
$T\sim{J}S^2$ which is deep inside the 'quantum critical' regime for
large value of $S$. For small value of $S$ where there is no
clear separation between energy scales $SJ$ and $S^2J$, the
whole 'quantum critical' regime may be washed away by the existence
of topological spin excitations. Such scenario was indeed
seemed to be observed in Monte Carlo simulations of $S=1/2$
2D Heisenberg antiferromagnet\cite{mc1,mc2,mc3} where the 'quantum
critical' regime predicted by NL$\sigma$M description
seems to be missing. Notice that for integer spin systems
the contribution to dynamics structure factor $S(\vec{q},\omega)$ 
from topological spin excitations are quite different from  
half-integer spin systems because of different statistics. However
their effect on $\xi(T)$ should be qualitatively similar. 
  
Lastly we discuss the effects of gauge field $\vec{A}$.
In the $T\rightarrow0$ limit where bose-condensation
of boson spin takes place ($<Z>\neq0$) the gauge field
$\vec{A}$ becomes confining\cite{frakin} and the topological
spin excitations are confined in pairs by linear confining
potential. At finite temperatures the confining potential
is effective up to length scale $\sim$ correlation length
$\xi(T)$ and the effect of confinement is expected to be strong
at low temperature when the system is at the 'renormalized classical'
regime. As a result the topological spin excitation spectrum
at $\vec{q}\sim(0,0)$ is expected to be
strongly modified from the free Dirac fermion prediction
(for half-integer spins) at low temperature. However the
behaviour of topological spin excitation spectrum at $\vec{q}
\sim(\pi,\pi)$  is determined by short distance behaviours of
spin pairs and should not be affected strongly by confinement.
As a result, we expect that our predictions for $S(\vec{q},\omega)$
at $\vec{q}\sim(\pi,\pi)$ and $\xi(T)$ remains qualitatively valid.

  Summarizing, using SBMFT, we show in this paper that
stable topological spin excitations exist in Heisenberg
antiferromagnets at two dimensions. The topological spin
excitations are described by relativistic quantum
field theories of complex scalars when spin value $S$
is an integer, and are Dirac-fermions when $S$ is
half-integer. Theoretical and experimental consequences of these
excitations are discussed where we point out that the existence
of these spin excitations may be the reason for the anomalous
results observed in recent experiments\cite{e1} and Monte-Carlo
simulations\cite{mc1,mc2,mc3,mc4}.  In particular, these
excitations can be observed directly in neutron scattering
experiments, and would provide a direct test of our theory.

  The author thanks Prof. N. Nagaosa and Prof. P.A. Lee for many
interesting discussions and also thanks the hospitality of the
Aspen Center of Physics where part of this work is completed. This
work is support by HKRGC through Grant no. HKUST6143-97P.

\end{document}